\pdfoutput = 1

\documentclass[12pt,a4paper]{article}

\usepackage{cite}
\usepackage{placeins}

\usepackage[utf8]{inputenc}

\usepackage{amsmath}
\usepackage[dvipsnames]{xcolor}
\usepackage{amssymb}
\usepackage{graphicx}
\usepackage[colorlinks=true, allcolors=Blue]{hyperref}
\usepackage{caption}
\usepackage{subcaption}
\usepackage{braket}
\usepackage{slashed}
\usepackage{float}
\usepackage{multirow}
\usepackage{cite}

\mathcode`l="8000
\begingroup
\makeatletter
\lccode`\~=`\l
\DeclareMathSymbol{\lsb@l}{\mathalpha}{letters}{`l}
\lowercase{\gdef~{\ifnum\the\mathgroup=\m@ne \ell \else \lsb@l \fi}}%
\endgroup
\usepackage{xspace}

\newcommand{\Bbar}{\kern 0.18em\overline{\kern -0.18em B}{}\xspace}

\newcommand{\Kbar}{\kern 0.18em\overline{\kern -0.18em K}{}\xspace}

\begin{document}
\begin{titlepage}

\vspace*{-0.7truecm}

\vspace{1.6truecm}

\begin{center}
\boldmath
{\Large{\bf Revealing New Physics in \boldmath$B^0_s\to D_s^\mp K^\pm$ Decays}}
\unboldmath
\end{center}

\vspace{0.8truecm}

\begin{center}
{\bf Robert Fleischer\,${}^{a,b}$ and  Eleftheria Malami\,${}^{a}$}

\vspace{0.5truecm}

${}^a${\sl Nikhef, Science Park 105, NL-1098 XG Amsterdam, Netherlands}

${}^b${\sl  Department of Physics and Astronomy, Vrije Universiteit Amsterdam,\\
NL-1081 HV Amsterdam, Netherlands}

\end{center}

\vspace*{1.7cm}

\begin{abstract}
\noindent
The $B^0_s\to D_s^\mp K^\pm$ system offers a determination of the Unitarity Triangle angle $\gamma$. Intrigued by an LHCb analysis showing a surprisingly large result in tension with information on the Unitarity Triangle and other $\gamma$ measurements, we make a transparent study of the measured observables, confirming the LHCb picture. The corresponding $\gamma$ puzzle at the $3\sigma$ level would require CP-violating contributions of New Physics, which should also manifest themselves in the corresponding decay branching ratios. Indeed, we find that the rates of the individual $B^0_s\to D_s^\mp K^\pm$ channels show puzzling patterns, in accordance with similar decays, with tensions up to $4.8\sigma$, thereby making the situation much more exciting. We present a formalism to include New-Physics effects in a model-independent way and apply it to the data to constrain the corresponding parameters. Interestingly, new contributions of moderate size could accommodate the data. Utilising this formalism in the future high-precision $B$ physics era may allow us to finally establish new sources of CP violation.  
\end{abstract}

\vspace*{2.1truecm}

\vfill

\noindent
September 2022

\end{titlepage}

\newpage

\thispagestyle{empty}
\vbox{}
\newpage

\setcounter{page}{1}

\section{Introduction} 
Decays of $B$ mesons provide a wide spectrum of probes for testing the quark-flavour sector of the Standard Model (SM). A particularly 
interesting aspect is given by the subtle difference between weak interactions of particles and their antiparticles, which is described by
the phenomenon of CP violation. Here C stands for charge conjugation and P refers to parity, i.e.\ space inversion. In the SM, CP violation is described through a complex phase in the Cabibbo--Kobayashi--Maskawa (CKM) matrix. Since the SM cannot 
explain the matter--antimatter asymmetry of the Universe, failing by many order of magnitude, new sources of CP violation are suggested.

The decays $\bar{B}^0_s\to D_s^+K^-$ and $B^0_s\to D_s^+ K^-$  (with their CP conjugates) offer a powerful probe for testing the 
SM description of CP violation \cite{ADK,RF-BsDsK,DeBFKMST}. Due to quantum-mechanical $B^0_s$--$\bar{B^0_s}$ oscillations, interference effects are generated between these decay channels, leading to the following time-dependent rate asymmetry: 
\begin{equation} \label{CP-asym}
	 \frac{\Gamma(B^0_s(t)\to D_s^{+} K^-) - \Gamma(\bar{B}^0_s(t)\to D_s^{+} K^-) }
	{\Gamma(B^0_s(t)\to D_s^{+} K^-) + \Gamma(\bar{B}^0_s(t)\to D_s^{+} K^-) }  
	 = \frac{{C}\,\cos(\Delta M_s\,t) + {S}\,\sin(\Delta M_s\,t)}
	{\cosh(y_s\,t/\tau_{B_s}) + {\cal A}_{\Delta\Gamma}\,\sinh(y_s\,t/\tau_{B_s})};
\end{equation}
an analogous expression holds for the CP-conjugate $D_s^{-} K^+$ final state, where $C$, 
$S$ and ${\cal A}_{\Delta\Gamma}$ are replaced by $\overline{C}$, $ \overline{S}$
and $\overline{{\cal A}}_{\Delta\Gamma}$, respectively. 

The $\bar{B}^0_s\to D_s^+K^-$ and $B^0_s\to D_s^+ K^-$ channels arise from $b\to c \bar{u}s$ and $\bar{b}\to \bar{u} c \bar{s}$ quark-level processes, respectively. In the SM, the observables of Eq.\ (\ref{CP-asym}) and its CP conjugate allow a theoretically clean 
determination of the angle $\gamma$ of the Unitarity Triangle (UT) of the CKM matrix \cite{ADK,RF-BsDsK,DeBFKMST}. Performing these measurements and assuming certain SM relations, LHCb \cite{BsDsK-LHCb-CP} has reported the result 
$\gamma = \left(128 ^{+17}_{-22} \right)^\circ$
modulo $180^\circ$. This finding is puzzling as global SM analyses of the UT give values around $70^\circ$ 
\cite{Amhis:2019ckw,PDG,CKMfitter,UTfit}. This regime is also consistent with a recent simultaneous analysis of various $B$-meson decays by the LHCb collaboration \cite{LHCb:2021dcr}. However, these modes have dynamics different from the $B^0_s\to D_s^\mp K^\pm$ channels and show sensitivity on $\gamma$ through very different interference effects. 

Intrigued by the challenge of interpreting this result, we have a transparent look at the determination of $\gamma$ and the associated
parameters, confirming the LHCb picture and excluding the modulo $180^\circ$ ambiguity \cite{Fleischer:2021cct}. If confirmed through future more precise 
measurements, this result could only be explained through CP-violating contributions of New Physics (NP) at the $B^0_s\to D_s^\mp K^\pm$ decay amplitude level, and should then also leave imprints on the corresponding branching ratios. Consequently,  we move on to extract the individual 
$\bar{B}^0_s\to D_s^+K^-$ and $B^0_s\to D_s^+ K^-$ branching ratios and compare them with the SM picture, developing a method to minimise 
dependences on non-perturbative hadronic and CKM parameters. We in fact arrive at yet another puzzling situation, complementing the $\gamma$ measurement. The situation is even more exciting, as we observe similar patterns also in branching ratios for other $B$ decays with similar dynamics.
Surprisingly small branching ratios in the latter modes have already received attention \cite{FST-BR,Bordone:2020gao}, also within NP interpretations \cite{Iguro:2020ndk,Cai:2021mlt,Bordone:2021cca}.  

In view of these findings, we present a generalized description of the $B^0_s\to D_s^\mp K^\pm$ decays and corresponding analysis of CP violation to 
include NP contributions in a model-independent way. Applying this formalism to the current data, we calculate correlations between the NP parameters, involving in particular new sources of CP violation.  This compact paper complements a much more detailed discussion given in Ref.~\cite{Fleischer:2021cct}.

\section{Standard Model Framework}\label{sec:SM}
The interference effects between $B^0_s$ or $\bar{B}^0_s$ mesons decaying into the final states $D_s^+K^-$ and $D_s^-K^+$ caused by the 
$B^0_s$--$\bar{B}^0_s$ oscillations are described by physical observables $\xi$ and $\bar\xi$, respectively. In the SM, their product provides 
a theoretically clean expression \cite{RF-BsDsK}:
\begin{equation} \label{multxi}
 {\xi} \times \bar{\xi}= e^{-i2( \phi_s + \gamma)},
\end{equation} %
where $\phi_s$ is the CP-violating $B^0_s$--$\bar{B}^0_s$ mixing phase. It should be noted in particular that non-perturbative hadronic matrix 
elements cancel in (\ref{multxi}). Since $\xi$ and  $\bar\xi$ can be determined from 
$C$, $S$, ${\cal A}_{\Delta\Gamma}$ and $\overline{C}$, $ \overline{S}$, 
$\overline{{\cal A}}_{\Delta\Gamma}$, respectively \cite{DeBFKMST}, and $\phi_s$ through measurements of CP violation in 
$B^0_s \rightarrow J/\psi \phi$ \cite{Barel:2020jvf}, this relation allows a clean extraction of $\gamma$. 

Furthermore, the following relations hold  in the SM framework \cite{RF-BsDsK}:
\begin{equation}\label{xi-rel}
| \bar{\xi}|={1}/{|\xi|} = \sqrt{(1+C)/{(1-C)}}, \ \ \ \ \ C+\overline{C}=0,
 \end{equation}
which were assumed by the LHCb collaboration \cite{BsDsK-LHCb-CP}. Performing a sophisticated fit to their data, LHCb found  
\begin{equation}\label{xi-L}
|\bar{\xi}|=0.37^{+0.10}_{-0.09}
\end{equation}
 with the following results \cite{BsDsK-LHCb-CP}:
\begin{equation}\label{LHCb-par-res}
\phi_s +\gamma= \left(126^{+17}_{-22}\right)^\circ, \ \ \delta_s= (-2^{+13}_{-14})^\circ \  \ {\text{[modulo $180^\circ$],}}
\end{equation}
were $\delta_s$ describes the CP-conserving strong phase difference between the $\bar{B}^0_s\to D_s^+K^-$ and 
$B^0_s\to D_s^+ K^-$ amplitudes. 
Here we have used $\phi_s=(-1.7\pm1.9)^\circ$, which was employed in the LHCb analysis, to convert $\gamma$ into $\phi_s+\gamma$. 
Using the updated result $\phi_s=\left(-5^{+1.6}_{-1.5}\right)^\circ$, which includes also penguin corrections in $B^0_s\to J/\psi \phi$ modes \cite{Barel:2020jvf},  we obtain 
\begin{equation}\label{gamma-res-1}
\gamma=\left(131^{+17}_{-22}\right)^\circ .
\end{equation} 

In view of the tension of this value with the SM and the complex LHCb analysis, it is crucial to transparently understand the situation. How can we achieve that? Using (\ref{xi-rel}) with the measured value of $C$, we find $| \bar{\xi}|=0.40 \pm 0.13$, which is in excellent agreement with (\ref{xi-L}). 
Introducing the combinations
\begin{equation}
\langle S \rangle_\pm\equiv (\overline{S}\pm S)/2, \quad \langle \mathcal{A}_{\Delta \Gamma} \rangle_+\equiv(\overline{{\cal A}}_{\Delta\Gamma}+{\cal A}_{\Delta\Gamma})/2,
\end{equation}
we obtain the following relations \cite{RF-BsDsK,DeBFKMST,Fleischer:2021cct}:
\begin{equation}\label{tanPhi}
\tan(\phi_s+\gamma)=-\langle S \rangle_+/ \langle \mathcal{A}_{\Delta \Gamma} \rangle_+=-1.45^{+0.73}_{-2.76}
\end{equation}
\vspace*{-0.7truecm}
\begin{equation}\label{tan-del}
\tan\delta_s=\langle S \rangle_- /\langle \mathcal{A}_{\Delta \Gamma} \rangle_+=0.04^{+0.70}_{-0.40},
\end{equation}
where the numerical values correspond to the LHCb measurements of the corresponding observables, yielding $\phi_s +\gamma=(125^{+18}_{-22})^\circ$ 
with $\delta_s=(2^{+34}_{-22})^\circ$. Here we have excluded the solutions modulo $180^\circ$ as they would be in huge conflict with factorization 
predicting $\delta_s\sim0^\circ$ \cite{Fleischer:2021cct}. Consequently, we find excellent agreement between this simple -- but transparent -- analysis and the complex LHCb fit.

The result for $\gamma$ in (\ref{gamma-res-1}) is much larger than the $70^\circ$ regime, and shows a discrepancy at the $3\sigma$ level. It is important to stress that this intriguing tension could not come from any long distance effects, 
as the $\gamma$ determination is theoretically clean in SM. Should this puzzle remain once the experimental picture sharpens further, it would require
new sources of CP violation. Since the experimental value of $\phi_s$ used in the analysis includes possible CP-violating NP effects in 
$B^0_s$--$\bar{B}^0_s$ mixing, new contributions entering directly at the decay amplitudes of the $B^0_s\to D_s^\mp K^\pm$ system would be required. 

Such NP effects should manifest themselves also in the corresponding branching ratios. Let us therefore focus on these observables next. Due to the time-dependent $B^0_s$--$\bar{B}^0_s$ oscillations, we have to distinguish between time-integrated ``experimental" branching ratios \cite{Dunietz:2000cr} and their ``theoretical" counterparts where such mixing effects are ``switched off" \cite{DeBFKMST,DeBruyn:2012wj}. Moreover, we have to disentangle the interference effects between the two decay paths that arise from the $B^0_s$--$\bar{B}^0_s$ oscillations. We obtain the following expressions for the theoretical branching ratios \cite{Fleischer:2021cct}:
\begin{align}\label{BR-Ds+K-}
\mathcal{B}(B^0_s\to D_s^+K^-)_{\text{th}}&=2 \left(\frac{1}{1+|\xi|^2} \right)\mathcal{B}_{\text{th}} \\
\mathcal{B}(\bar B^0_s\to D_s^+K^-)_{\text{th}}&=
2 \left(\frac{ |\xi|^2 }{1+|\xi|^2} \right)\mathcal{B}_{\text{th}}= |\xi|^2 \mathcal{B}(B^0_s\to D_s^+K^-)_{\text{th}}.
\end{align}
Analogous expressions hold for the branching ratios of the $B^0_s$ and $\bar{B}^0_s$ decays into the final state $D_s^- K^+$. Unfortunately, separate measurements of the experimental branching ratios for these final states are not available, just the average
\begin{equation}
\langle\mathcal{B}_{\text{exp}}\rangle\equiv \frac{1}{2}\left(\mathcal{B}_{\text{exp}} + \bar{\mathcal{B}}_{\text{exp}}\right)
\equiv\frac{1}{2} \, \mathcal{B}^{\text{exp}}_\Sigma
\end{equation}
with $\mathcal{B}^{\text{exp}}_\Sigma= (2.27 \pm 0.19) \times 10^{-4}$ \cite{PDG}. 
Assuming SM expressions for the decay amplitudes, we have  \cite{RF-BsDsK,DeBFKMST}
\begin{equation}\label{SM-BR-3}
 \mathcal{B}_{\text{th}} =  \bar{\mathcal{B}}_{\text{th}}=
\left[\frac{1-y_s^2}{1+y_s\langle {\cal A}_{\Delta\Gamma}\rangle_+}\right]\langle\mathcal{B}_{\text{exp}}\rangle
\end{equation}
with $y_s=0.062 \pm 0.004$ \cite{PDG} and 
$\langle \mathcal{A}_{\Delta \Gamma} \rangle_+= 
0.35 \pm 0.23$ \cite{Fleischer:2021cct}.
Finally, we obtain 
\cite{Fleischer:2021cct}:
\begin{align}
 \mathcal{B}(B^0_s \rightarrow D_s^{+}K^{-})_{\rm th}&=(0.26 \pm 0.12) \times 10^{-4} \label{BR-2} \\
\mathcal{B}(\bar{B}^0_s \rightarrow D_s^{+}K^{-})_{\rm th}&=(1.94 \pm 0.21) \times 10^{-4} \label{BR-1}.
\end{align}

For the calculation of these branching ratios and the underlying decay amplitudes, factorization provides the theoretical framework. 
Here the corresponding hadronic matrix elements of four-quark operators entering the low-energy effective Hamiltonian 
are factorized into the product of the matrix elements of quark currents. 

The decay ${\bar{B}^0_s \rightarrow D_s^+ K^-}$ originating from $b\to c \bar u s$ processes is a prime example where 
``QCD factorization" is expected to work excellently for the colour-allowed tree topologies \cite{Beneke:2000ry,bjor,DG,Neubert:1997uc,SCET}.
We obtain the following amplitude:
\begin{equation}\label{eq:a1DsK}
A_{D_s^+ K^-}^{\text{SM}} = \frac{G_{\rm F}}{\sqrt{2}} V_{us}^{\ast}V_{cb} f_K  F_0^{B_s \rightarrow D_s}(m_K^2)  (m_{B_s}^2 - m_{D_s}^2)  a_{\rm 1 \, eff }^{D_s K},
\end{equation}
where $G_{\rm F}$ is the Fermi constant, $V_{us}^{\ast}V_{cb}$ a factor of CKM matrix elements, $f_K$ the kaon decay constant and $F_0^{B_s \rightarrow D_s}(m_K^2)$ a form factor parametrising the hadronic $b \rightarrow c$ 
quark-current matrix element. The parameter 
\begin{equation}\label{a-eff-1-DsK}
a_{\rm 1 \, eff }^{D_s K}=a_{1}^{D_s K} \left(1+\frac{E_{D_s K}}{T_{D_s K}}\right) 
\end{equation}
describes the deviation from naive factorization: $a_{1}^{D_s K}$ characterises non-factorisable effects entering the colour-allowed tree amplitude $T_{D_s K}$, whereas $E_{D_s K}$ denotes non-factorisable exchange topologies. 

The current state-of-the-art results 
within QCD factorization are found as $|a_1|~\approx~1.07$ with a quasi-universal behaviour \cite{Bordone:2020gao,Huber:2016xod}, with uncertainties at the percent level. Recently, even QED effects have been studied \cite{Beneke:2021jhp}, which are small and fully included within the uncertainties.In Ref.~\cite{Bordone:2020gao}, a theoretical analysis of the $\bar{B}^0_d\to D^+K^-$ decay, which does not have an exchange topology, has been performed, 
yielding $|a_1^{D_dK}| = 1.0702^{+0.0101}_{-0.0128}$. The same numerical result is found for $\bar B^0_s\to D_s^{+}\pi^-$, yet another pure colour-allowed tree decay which does not have an exchange contribution. A detailed discussion of the associated uncertainties is also given in \cite{Bordone:2020gao}. The $\bar{B}^0_s\to D_s^+K^-$ channel -- the key player for our analysis -- differs from the $\bar{B}^0_d\to D^+K^-$ mode 
only through the spectator quark. 
We use the $SU(3)$ flavour symmetry to relate the spectator quarks to each other and double the tiny error in view of possible $SU(3)$-breaking effects, 
and employ the following value for our analysis of the $b\to c \bar u s$  transition: 
\begin{equation}\label{a1-1-pred}
|a_1^{D_sK}| = 1.07\pm0.02.
\end{equation}

The exchange topology in the $\bar{B}^0_s\to D_s^+K^-$ decay is non-factorizable and cannot be reliably calculated from 
first principles. Consequently, we use experimental data to constrain this contribution. The $\bar{B^0_d} \to D_s^+ K^-$ transition
originates only from an exchange topology $E_{D_sK}'$, which differs from its counterpart $E_{D_sK}$ in $\bar B^0_s \to D_s^+K^-$ 
through the down quark of the initial $\bar B^0_d$ meson. To be specific, we have the expression
\begin{equation}
\hspace*{-0.2truecm}\left|\frac{E_{D_sK}'}{T_{D_sK}+E_{D_sK}}\right|^2=
\frac{\tau_{B_s}}{\tau_{B_d}}\frac{m_{B_d}}{m_{B_s}}
\left[\frac{\Phi(m_{D_s}/m_{B_s}, m_{K}/m_{B_s})}{\Phi(m_{D_s}/m_{B_d}, m_{K}/m_{B_d})}\right]
\left|\frac{V_{us}}{V_{ud}}\right|^2
\left[\frac{\mathcal{B}(\bar{B^0_d} \to D_s^+ K^-)}{\mathcal{B}(\bar{B^0_s} \to D_s^+ K^-)_{\rm th}}\right],
\end{equation}
where $\tau_{B_{d}}$ and $\tau_{B_s}$ denote the lifetimes of the $B_d$ and $B_s$ mesons, respectively, and $\Phi(x,y)$ is the usual phase-space function, depending on the meson masses. Using this expression with the measured branching ratio
${\cal B}(\bar{B^0_d} \to D_s^+ K^-)=(2.7\pm0.5)\times10^{-5}$ and our result in (\ref{BR-1}), we find 
\begin{equation}\label{E-rat-1}
\left|\frac{E_{D_sK}'}{T_{D_sK}+E_{D_sK}}\right| = 0.08 \pm 0.01,
\end{equation}
which offers direct access to the size of the exchange contribution. Another constraint is provided through the comparison of the branching ratio of 
$\bar{B}^0_s\to D_s^+K^-$ in (\ref{BR-1}) with ${\cal B}(\bar{B}^0_d\to D^+K^-)=(1.86\pm0.20)\times10^{-4}$, yielding the following result
 \cite{Fleischer:2021cct}:
\begin{equation}\label{rE} 
 r_E^{D_sK} \equiv \left|1+\frac{E_{D_s K}}{T_{D_s K}}\right|=1.00\pm0.08,
\end{equation}
which is remarkably consistent with (\ref{E-rat-1}). Due to the non-factorizable nature of the exchange amplitude, it may well have a large strong phase 
difference with respect to the colour-allowed tree amplitudes \cite{DeBFKMST,FST-BR}. Interestingly, this feature is indicated by the comparison of 
(\ref{E-rat-1}) with (\ref{rE}), although the current uncertainties do not allow us to draw further conclusions. It is important to emphasize that no anomalous behaviour of the exchange topologies that could be caused by large rescattering or other non-factorizable effects is indicated by the data \cite{Fleischer:2021cct}, as was also found in Ref.~\cite{FST-BR}.

\begin{figure}[t!]
	\centering
\hspace{-0.1cm}\includegraphics[width = 0.48\linewidth]{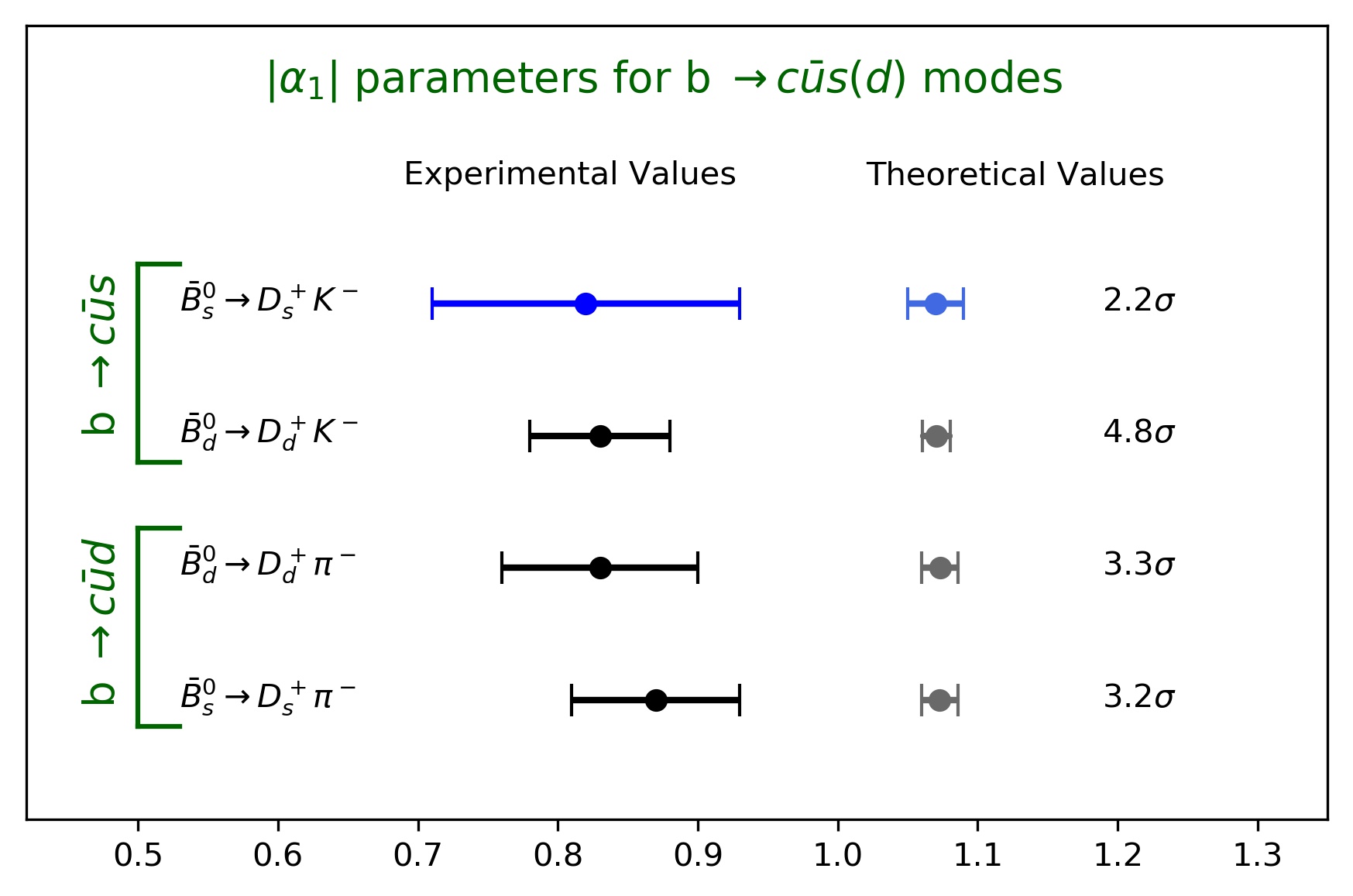}
\includegraphics[width = 0.48\linewidth]{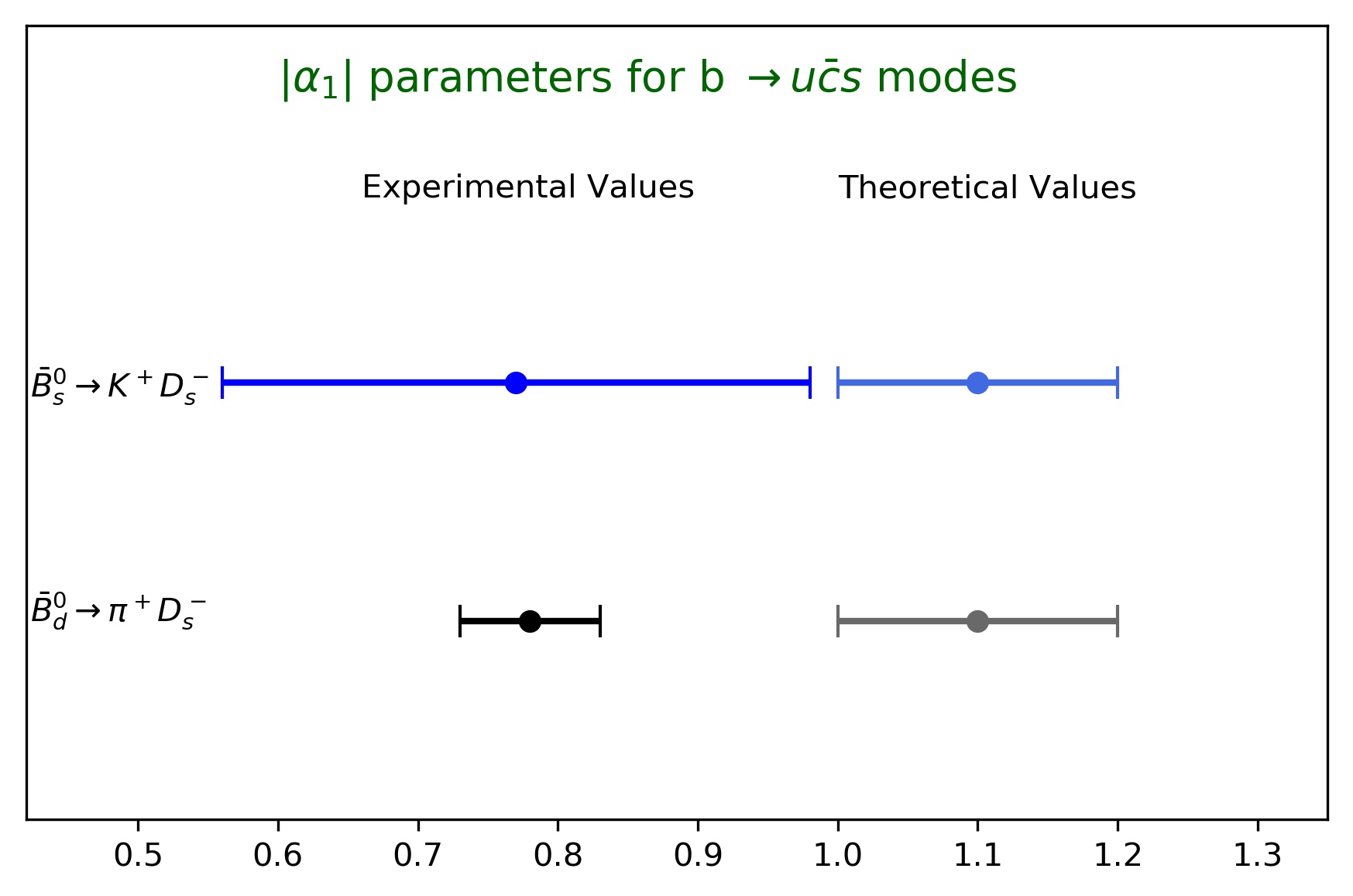}
\vspace*{-0.3truecm}
	\caption{Experimental and theoretical SM values of the $|a_1|$ parameters for various decay processes. The left panel illustrates decays which are caused by $b\to c \bar u s$ and $b\to c \bar u d$ processes while the right panel shows decays originating from $b\to u\bar c s$ transitions.} \label{fig:aval}
\end{figure}

Our next step is to extract the $|a_1|$ parameters from the data in the cleanest possible way and to compare them with the
theoretical SM predictions. In this respect, semileptonic decays provide a very useful tool 
\cite{Beneke:2000ry,FST-BR,Neubert:1997uc}.
In the case of the $\bar{B}^0_s\to D_s^+K^-$ channel, we have the partner decay 
$\bar{B}^0_s \rightarrow D_s^{+}\ell^{-} \bar{\nu}_{\ell}$, and introduce the ratio
\begin{equation}
  R_{D_s^{+}K^{-}}\equiv\frac{\mathcal{B}(\bar{B}^0_s \rightarrow D_s^{+}K^{-})_{\rm th}}{{\mathrm{d}\mathcal{B}\left(\bar{B}^0_s \rightarrow D_s^{+}\ell^{-} \bar{\nu}_{\ell} \right)/{\mathrm{d}q^2}}|_{q^2=m_{K}^2}},
    \label{semi}
\end{equation}
which takes the form
\begin{equation}\label{RDsKm-expr}
R_{D_s^{+}K^{-}}=6 \pi^2 f_{K}^2 |V_{us}|^2 |a_{\rm 1 \, eff }^{D_s K}|^2  X_{D_s K}
\end{equation}
\vspace*{-0.1truecm}
with
\vspace*{-0.1truecm}
\begin{equation}  \label{RSM}
    X_{D_s K}\equiv{\Phi_{\text{ph}}} \left[ {F_0^{B_s \rightarrow D_s}(m_K^2)}/{F_1^{B_s \rightarrow D_s}(m_K^2)} \right]^2,
\end{equation}
where $\Phi_{\text{ph}}$ is a phase-space factor which is equal to 1 with excellent precision \cite{Fleischer:2021cct}. The CKM matrix element $|V_{cb}|$ cancels 
in $R_{D_s^{+}K^{-}}$. Moreover, due to the normalization condition $F_0^{B_s \rightarrow D_s}(0)=F_1^{B_s \rightarrow D_s}(0)$, we 
have an essentially negligible impact of the non-perturbative hadronic form factors \cite{Monahan:2017uby,McLean:2019qcx,Aoki:2019cca}. Using a recent LHCb measurement \cite{LHCb:2020cyw}
of the differential rate of the
$\bar{B}^0_s \rightarrow D_s^{+}\ell^{-} \bar{\nu}_{\ell}$ channel and (\ref{BR-1}) with (\ref{rE}) yields
\begin{equation}\label{a-DsK-exp}
|a_{\rm 1}^{D_s K}| = 0.82 \pm 0.11.
\end{equation}
We observe that this result is in tension with the theoretical prediction in (\ref{a1-1-pred}) at the $2.2\,\sigma$ level. Consequently, we indeed find another tension at the decay amplitude level with respect to the SM, as we would expect in view of the puzzling result for $\gamma$. This exciting situation is 
further strengthened by the fact that a similar pattern of the $|a_1|$ parameters -- with surprisingly small values -- arises also for other $B_{(s)}$ decays 
with similar dynamics. We have extracted these quantities from the data in an analogous way with semileptonic decay information \cite{Fleischer:2021cct}, 
and show the results in the left panel of Fig.~\ref{fig:aval}. Here $\bar B^0_d\to D_d^+ K^-$ stands out, showing even a discrepancy of $4.8 \,\sigma$. 
Puzzlingly small 
branching ratios for this channel and the $\bar B^0_d\to D_d^+ \pi^-$, $\bar B^0_s\to D_s^+ \pi^-$ modes were also pointed out in 
Refs.\ \cite{FST-BR,Bordone:2020gao}. 

Let us next have a look at the $\bar{B}^0_s \rightarrow K^+ D_s^-$ decay. The amplitude of this channel, which is caused by $b\to u\bar c s$ processes, 
can be expressed in a way similar to the $b\to c\bar u s$ case, as we discuss in detail in \cite{Fleischer:2021cct}. In analogy to the 
${\bar{B}^0_s \rightarrow D_s^+ K^-}$ mode, this transition is also a colour-allowed tree decay. However, as the roles of the heavy $c$ and light $u$ quarks are interchanged, the heavy-quark arguments which can be used to prove factorization up to tiny corrections for the $b\to c$ modes do not apply, and there may be larger non-factorizable effects. In view
of this less favourable theoretical situation, we will use the following range as a reference for our analysis:
\begin{equation}\label{a1-2-pred}
|a_1^{K D_s}| = 1.1 \pm 0.1.
\end{equation} 
Here we have followed the QCD renormalisation group analysis in Ref.~\cite{Buras:1994ij} as guidance, where $|a_1|=1.01\pm0.02$ was 
found for the variation of the global $|a_1|$ parameter for colour-allowed tree decays with respect to variations of the renormalization scale and schemes. 
In Eq.~(\ref{a1-2-pred}), we allow for a five times larger uncertainty \cite{Fleischer:2021cct}.
Interestingly, the experimental value of $\delta_s$ in (\ref{LHCb-par-res}) is found in excellent agreement with factorization \cite{RF-BsDsK}. Since this strong
phase difference characterizes the interference between the $b\to c\bar u s$ and $b\to u\bar c s$ decay paths, it supports factorisation -- where such phases vanish -- also in the $b\to u\bar c s$ channel. 

In analogy Eq.~(\ref{semi}), the partner decay for the clean extraction of the $|a_1^{K D_s}|$ parameter is $\bar B^0_s \rightarrow K^+ \ell \bar{\nu}_{\ell}$. Although this channel has been observed by LHCb \cite{LHCb:2020ist}, a measurement of the differential rate has not yet been reported. 
Consequently, we have applied the $SU(3)$ flavor symmetry and have utilised 
the $\bar B^0_d \rightarrow \pi^+ \ell^{-} {\bar{\nu}_{\ell}}$ mode, 
for which we do have information from the BaBar and Belle collaborations \cite{Amhis:2019ckw,PDG}. We find
\begin{equation}\label{a1KDs-extr}
|a_{\rm 1}^{K D_s}| =0.77 \pm 0.21,
\end{equation}
as discussed in more detail in \cite{Fleischer:2021cct}. As in the $b\to c \bar u s$ case, this result  favours  again a value smaller than our 
theoretical reference in Eq.\ (\ref{a1-2-pred}). The result is illustrated in the panel on the right-hand side of Fig.~\ref{fig:aval}, where we include also 
the $\bar{B}^0_d\to \pi^+D_s^-$ mode, which differs only through the spectator quark from the $\bar{B}^0_s \rightarrow K^+ D_s^-$ channel. The current uncertainties are too large to draw further conclusions on these modes.

Within the SM, universal power-suppressed corrections of order $\Lambda_{\rm QCD}/m_b$ could in principle lead to a suppression of the $|a_1|$ parameters \cite{Huber:2016xod}. However, such effects would not allow us to accommodate the puzzling result for $\gamma$ arising from the CP-violating 
observables of the $B^0_s\to D_s^\mp K^\pm$ system. This phenomenon would require new sources for CP violation. The exciting possibility of 
NP effects in non-leptonic tree-level decays of $B$ mesons was discussed in Refs.\ \cite{Brod:2014bfa,Lenz:2019lvd}, and 
models for physics beyond the SM addressing the puzzles of the small branching ratios were studied in Refs.\ \cite{Iguro:2020ndk,Cai:2021mlt}. 
We consider these first models, which face challenges from direct NP searches at ATLAS and CMS \cite{Bordone:2021cca}, as interesting 
illustrations of specific scenarios. In the remainder of this paper, we will work with a model-independent parametrization of NP contributions.

\section{New Physics Analysis}

Let us now extend the analysis of the $B^0_s\to D_s^\mp K^\pm$ system to include NP effects. As we have already noted, the NP contributions to accommodate the puzzling measurement of $\gamma$ and the $|a_1|$ values would have to 
enter at the decay amplitude level. 
For the  $b\to c \bar u s$ quark-level transition, we generalise the decay amplitude as 
\begin{equation}
A(\bar{B}^0_s \rightarrow D_s^+ K^-) = A(\bar{B}^0_s \rightarrow D_s^+ K^-)_{{\text{SM}}} \left[ 1 + \bar{\rho} \, e^{i \bar{\delta}}
e^{+i \bar{\varphi}} \right]
\end{equation}
with the NP parameter
\begin{equation}
\bar{\rho} \, e^{i \bar{\delta}} e^{i \bar{\varphi}}  \equiv \frac{ A(\bar{B}^0_s \rightarrow D_s^+ K^-)_{{\text{NP}}} }{  A(\bar{B}^0_s \rightarrow D_s^+ K^-)_{{\text{SM}}} }, 
\end{equation}
where $\bar{\varphi}$ and $ \bar{\delta}$ denote CP-violating and CP-conserving phases, respectively. A similar expression can be written for the $\bar b \to \bar u c \bar s$ transition, with parameters $\rho$, $\varphi$ and $\delta$.

In order to get first access to these NP pazs, we compare the SM predictions for the branching ratios with the corresponding experimental values. Let us first have a look at the $b\to c \bar u s$ transition, where we introduce 
\begin{equation}\label{b-bar-def1}
\bar b \equiv  \frac{\langle\mathcal{B}(\bar {B}^0_s \rightarrow D_s^+K^-)_{\rm th}  \rangle}{\mathcal{B}(\bar{B}^0_s \rightarrow D_s^{+}K^{-})_{\rm th}^{\rm SM}}  = 1+2 \, \bar\rho\cos\bar\delta\cos\bar\varphi + \bar\rho^2.
\end{equation}
Here $\langle\mathcal{B}(\bar {B}^0_s \rightarrow D_s^+K^-)_{\rm th}  \rangle$ is the CP average of the theoretical branching ratios, converted from the experimental measurements. For the determination of $\bar b$, we utilize again the semileptonic ratio in (\ref{semi}), which is particularly clean with respect to 
form factor and CKM parameter uncertainties. In the presence of NP contributions with new sources of CP violation, 
we introduce the following generalized ratio
 \cite{Fleischer:2021cct}:
\begin{equation}\label{eq:R}
 \langle R_{D_s K} \rangle \equiv 
 \frac{\mathcal{B}(\bar{B}^0_s \rightarrow D_s^{+}K^{-})_{\rm th} + 
 \mathcal{B}({B}^0_s \rightarrow D_s^{-}K^{+})_{\rm th}}{\left[{\mathrm{d}\mathcal{B}\left(\bar{B}^0_s \rightarrow D_s^{+}\ell^{-} \bar{\nu}_{\ell} \right)/{\mathrm{d}q^2}}+ {\mathrm{d}\mathcal{B}\left({B}^0_s \rightarrow D_s^{-}\ell^{+} {\nu}_{\ell} \right)/{\mathrm{d}q^2}}\right]|_{q^2=m_{K}^2}} ,
\end{equation}
which satisfies $ \langle R_{D_s K } \rangle = R_{D_s^+ K^-}$ in the case of vanishing direct CP asymmetries, as in the SM.
Then we obtain the following expression for $\bar b $ in terms of $\langle R_{D_s K} \rangle$:
\begin{equation}\label{b-bar-def2}
\bar b   =\frac{\langle R_{D_s K} \rangle}{6 \pi^2 f_{K}^2 |V_{us}|^2 |a_{\rm 1 \, eff}^{D_s K}|^2 X_{D_s K} } .
\end{equation}
The parameter $|a_{\rm 1 \, eff}^{D_s K}|$ given in (\ref{a-eff-1-DsK}) is the product of the theoretical prediction of $|a_{1}^{D_s K}|$ in 
(\ref{a1-1-pred}) obtained within QCD factorization and the parameter $ r_E^{D_sK} $ in (\ref{rE}), which was constrained through experimental data. We obtain the value
\begin{equation} \label{eq:a1th}
|a_{\rm 1 \, eff}^{D_s K}| = 1.07 \pm 0.09,
\end{equation}
which we will use in the numerical analysis below. The small impact of the exchange topology following from the experimental data as reflected by Eq.~(\ref{E-rat-1}) holds irrespectively of whether we have the SM or possible NP contributions to this topology.

In analogy, for the $\bar b \to \bar u c \bar s$ transition, we introduce
\begin{equation}\label{b-def}
b\equiv 
1+2 \, \rho\cos\delta\cos\varphi + \rho^2 = \frac{\langle R_{K D_s}\rangle}{6 \pi^2 f_{D_s}^2 |V_{cs}|^2 |a_{\rm 1 \, eff}^{K D_s}|^2 X_{K D_s}} .
\end{equation}  
Making use of $r_E^{KD_s}=1.00\pm0.08$, which follows from an analysis similar to the one for $ r_E^{D_sK}$ given above  \cite{Fleischer:2021cct}, and the reference for $|a_{\rm 1}^{K D_s}|$ in 
Eq.~(\ref{a1-2-pred}), we find
\begin{equation} \label{eq:a1th-1}
|a_{\rm 1 \, eff}^{K D_s}|=1.1 \pm 0.13.
\end{equation} 

For the observable  $\bar{\xi} $, we obtain the generalisation 
\begin{equation}
\bar{\xi} =  \bar{\xi}_{\text{SM}}
\left[\frac{1 + {\rho} \, e^{i {\delta}} e^{+i {\varphi}}}{1 + \bar{\rho} \, e^{i \bar{\delta}} 
e^{-i \bar{\varphi}}}\right]
=-|\bar{\xi}|e^{+i\delta_s}e^{-i(\phi_s+\gamma)}e^{i\Delta\bar{\varphi}}.
\end{equation}
Similarly, we rewrite $\xi$ with $\Delta{\varphi}$, where we interchange the NP parameters $\bar{\rho}$, $\bar{\delta}$, $\bar{\varphi}$ and ${\rho}$, ${\delta}$, ${\varphi}$.

The product $\xi \times \bar{\xi} $ introduced in Eq.\ (\ref{multxi}), which plays the key role for CP violation in the SM, is generalized as follows:
\begin{equation}\label{xi-prod-NP}
\xi \times \bar{\xi}  
= e^{-i2 (\phi_s + \gamma)}
\Biggl[\frac{1 + {\rho} \, e^{i {\delta}} e^{+i {\varphi}}}{1 + {\rho} \, e^{i {\delta}} e^{-i {\varphi}}}
\Biggr]
\Biggl[\frac{1 + \bar{\rho} \, e^{i \bar{\delta}} e^{+i \bar{\varphi}}}{1 + \bar{\rho} \, e^{i \bar{\delta}} 
e^{-i \bar{\varphi}}} \Biggr].
\end{equation}
In contrast to the SM, NP may generate non-vanishing direct CP asymmetries:
\begin{equation}
{\cal A}^{\rm dir}_{\rm CP}\equiv \frac{|A({B}^0_s \rightarrow D_s^+ K^-)|^2-|A(\bar{B}^0_s \rightarrow D_s^- K^+)|^2}{|A({B}^0_s \rightarrow D_s^+ K^-)|^2+|A(\bar{B}^0_s \rightarrow D_s^- K^+)|^2} =\frac{2\,\rho\sin\delta\sin\varphi}{1+2 \, \rho\cos\delta\cos\varphi + \rho^2}.
\end{equation}
An analogous expression holds for the CP-conjugate asymmetry $\bar{{\cal A}}^{\rm dir}_{\rm CP}$, 
involving $\bar{\rho}$ with $\bar{\delta}$ and ${\bar\varphi}$.
We may then write the first ratio entering (\ref{xi-prod-NP}) as
\begin{equation}
\frac{1 + {\rho} \, e^{i {\delta}} e^{+i {\varphi}}}{1 + {\rho} \, e^{i {\delta}} 
e^{-i {\varphi}}} = e^{-i\Delta\Phi} \sqrt{\frac{1- {{\cal A}}^{\rm dir}_{\rm CP}}{1+{{\cal A}}^{\rm dir}_{\rm CP}}}, 
\end{equation}
where
\begin{equation}
\tan\Delta\Phi=-\left[\frac{2\rho\cos\delta\sin\varphi+
\rho^2\sin2\varphi}{1+2\rho\cos\delta\cos\varphi+
\rho^2\cos2\varphi}\right];
\end{equation}
the second ratio takes a similar form, involving $\bar\rho$, $\bar\varphi$ and $\bar{{\cal A}}^{\rm dir}_{\rm CP}$ with a phase
$\Delta\bar{\Phi}$.
We then obtain
\begin{equation}
\left|\xi\times\bar\xi\right|^2=
\left[ \frac{1- {{\cal A}}^{\rm dir}_{\rm CP}}{1+{{\cal A}}^{\rm dir}_{\rm CP}}\right]
\left[\frac{1- {\bar{\cal A}}^{\rm dir}_{\rm CP}}{1+{\bar{\cal A}}^{\rm dir}_{\rm CP}}\right]
=1+\epsilon,
\end{equation} 
where 
\begin{equation}
-\frac{1}{2}\, \epsilon = \frac{C+\bar{C}}{\left(1+C\right)\left(1+\bar{C}\right)} = 
{\cal A}^{\rm dir}_{\rm CP} + {\bar{\cal A}}^{\rm dir}_{\rm CP} + {\cal O}(({\cal A}^{\rm dir}_{\rm CP})^2),
\end{equation}
generalising the SM relations in (\ref{xi-rel}). Finally, we arrive at
\begin{equation}
\xi \times \bar{\xi}  = \sqrt{1-2\left[\frac{C+\bar{C}}{\left(1+C\right)\left(1+\bar{C}\right)}
\right]}e^{-i\left[2 (\phi_s +\gamma_{\rm eff})\right]},
\end{equation}
which is a theoretical clean relation playing key role in our analysis. Here,
the UT angle $\gamma$ enters as the ``effective" angle
\begin{equation}\label{gamma-eff}
\gamma_{\rm eff}\equiv 
 \gamma + \gamma_{\text{NP}}
 =\gamma+\frac{1}{2}\left(\Delta\Phi+\Delta\bar{\Phi}\right)=
\gamma-\frac{1}{2}\left(\Delta\varphi+\Delta\bar{\varphi}\right).
\end{equation}
Consequently, (\ref{gamma-res-1}) actually corresponds to $\gamma_{\rm eff}$.

Note that in combined fits to the data for various $B$ decays to extract $\gamma$, such as in Ref.~\cite{LHCb:2021dcr}, NP effects may 
average out to some extend, thereby yielding an effective angle with NP contributions which -- in contrast to (\ref{gamma-eff}) -- cannot transparently be quantified. It will rather be crucial to search for patterns in the individual $\gamma$ determinations, aiming at the highest precision.

Let us now apply our formalism to the current data. In order to be consistent with the LHCb assumption, we set the strong 
phases $\delta$ and $\bar{\delta}$ to $0^\circ$. 
This implies vanishing direct CP asymmetries, in agreement with $B \rightarrow DK$ data within the
uncertainties \cite{PDG}.As we have noted after Eq.~(\ref{eq:R}), using these assumptions, we may identify the experimental values for $R_{D_s^{+}K^{-}}=0.05 \pm 0.01$ and $R_{K^{+}D_s^{-}} = 3.64 \pm 1.70$, with $ \langle R_{D_s K } \rangle$ and $ \langle R_{K D_s  } \rangle$, respectively. Complementing these values with the other relevant parameters introduced above, we find
\begin{equation}\label{b-res-2} 
\bar b=0.58 \pm 0.16,  \qquad {b}=0.50 \pm 0.26.
\end{equation} 
Since the $\bar{b}$ and $b$ observables would be equal to 1 within the SM, these numerical values reflect the puzzling patterns in Fig.~\ref{fig:aval}. 
Concerning the input from CP violation, we obtain the \mbox{following relation \cite{Fleischer:2021cct}:}
\begin{equation}\label{gamma-res-0}
\Delta\varphi=\Delta\bar{\varphi} = \gamma-\gamma_{\rm eff}=-(61 \pm 20)^\circ,
\end{equation}
where 
\vspace*{-0.2truecm}
\begin{equation}\label{gamma-res-10}
\tan\Delta\varphi=\frac{\rho\sin\varphi+\bar{\rho}\sin\bar{\varphi}+\bar{\rho}\rho\sin(\bar{\varphi}+\varphi)}{1+\rho\cos\varphi +
\bar{\rho}\cos\bar{\varphi}+\bar{\rho}\rho\cos(\bar{\varphi}+\varphi)}.
\end{equation}
The numerical value refers to $\gamma=(70 \pm 7)^\circ$, which is consistent with UT analyses, 
and the result in (\ref{gamma-res-1}).

In order to convert the measured observables into constraints on the NP parameters, we first employ $\bar b$ and $b$ to
determine $\bar{\rho}$ and ${\rho}$ as functions of $\bar{\varphi}$ and $\varphi$, respectively. Using then (\ref{gamma-res-1}), 
we may calculate ${\varphi}$ as a function of $\bar\varphi$, fixing a contour in the
${\varphi}$--$\bar\varphi$ plane. Finally, using again $\bar{\rho}(\bar\varphi)$ and ${\rho}(\varphi)$ allows us to calculate
a correlation in the $\bar{\rho}$--$\rho$ plane, where each point is linked with $\bar\varphi$ and $\varphi$. 
\begin{figure}[t!]
	\centering
\hspace{-0.1cm}\includegraphics[width = 0.42\linewidth]{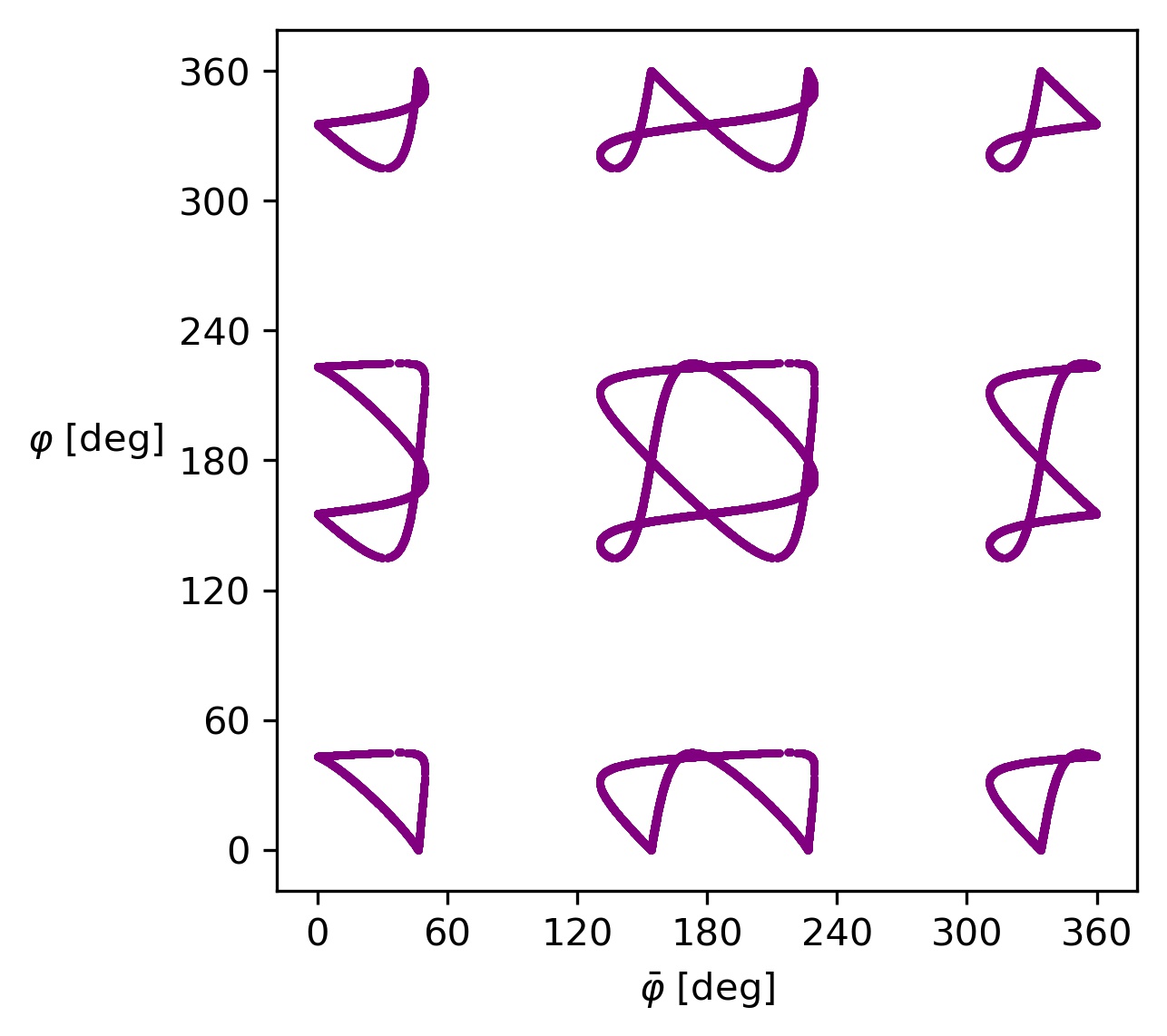} \hspace{0.5cm}
\includegraphics[width = 0.38\linewidth]{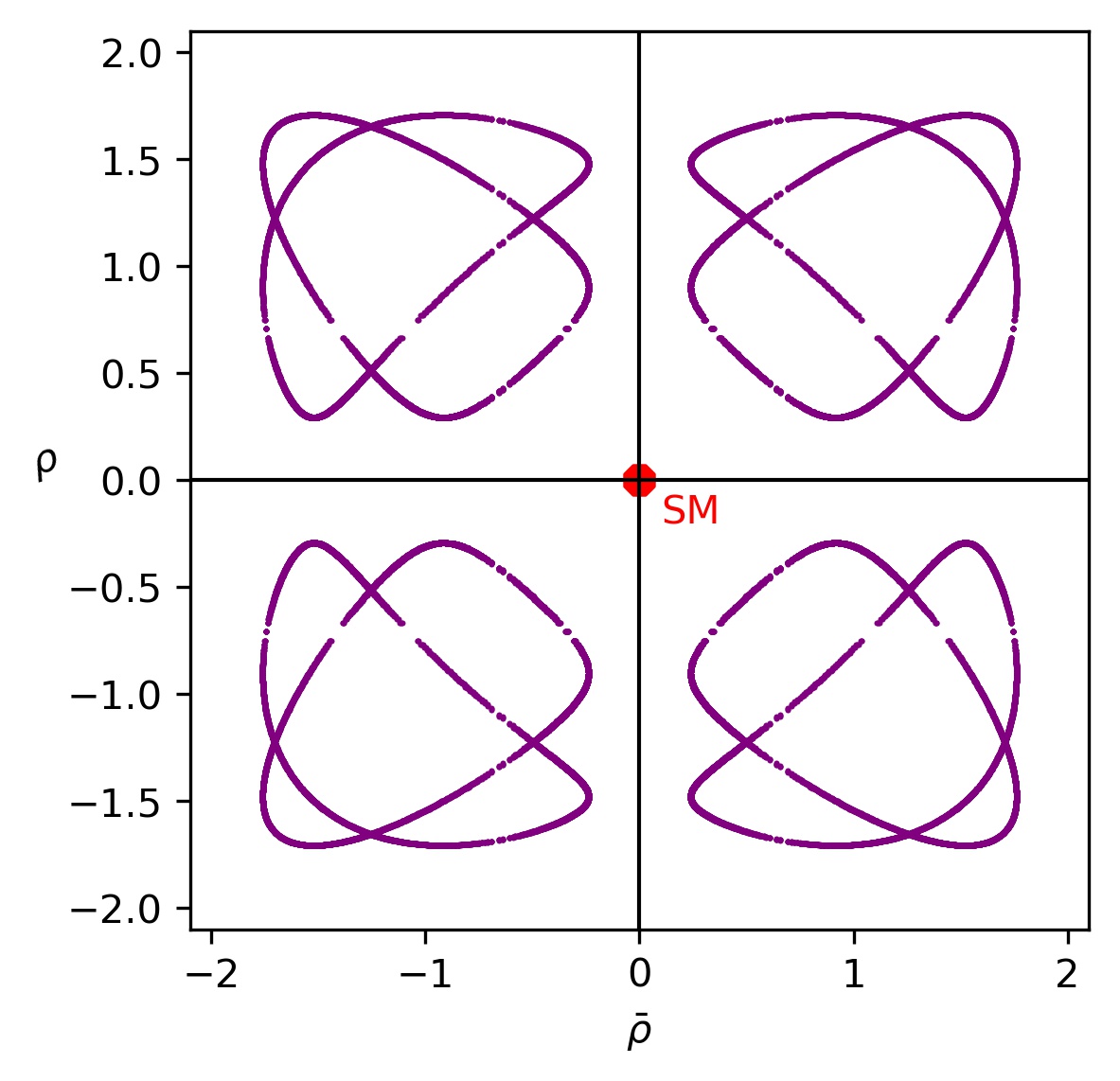}
\vspace*{-0.3truecm}
	\caption{Correlations for the central values of the current data in the $\bar{\varphi}$--$\varphi$ plane (left) and the $\bar{\rho}$--$\rho$ plane (right) of of NP parameters.}
	\label{fig:centrrhophi-1}
\end{figure}

\begin{figure}[t!]
	\centering
\includegraphics[width = 0.45\linewidth]{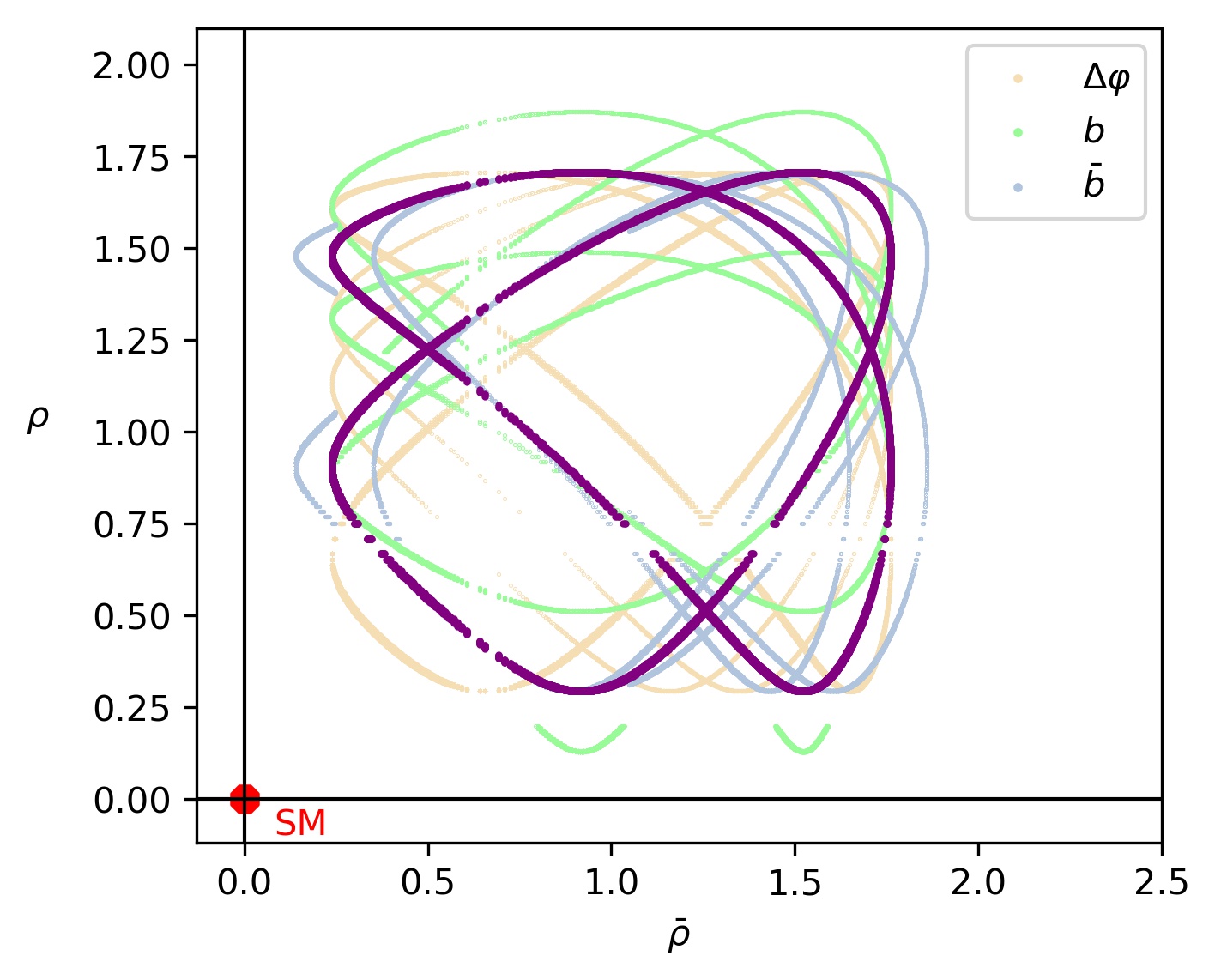}
\vspace*{-0.3truecm}
	\caption{Correlations in the $\bar{\rho}$--$\rho$ plane of NP parameters including uncertainties.}
	\label{fig:centrrhophier-1}
\end{figure}

In Fig.~\ref{fig:centrrhophi-1}, we show the corresponding correlations for the central values of the current data. 
In Fig.~\ref{fig:centrrhophier-1}, we show the impact of the uncertainties of the input quantities $\Delta\varphi$, $b$ and $\bar{b}$,
varying them separately. We can nicely see that the SM point corresponding to the origin in the $\bar{\rho}$--$\rho$ plane is excluded, and notice that NP contributions with CP-violating
phases are simultaneously required in the $b\to c \bar u s$ and $\bar b\to \bar u c \bar s$ decay paths. Interestingly, we can describe
 the measurements with NP contributions as small as about 30\% of the SM  amplitudes.
 
In view of the complexity of the strategy, we finally summarise the main steps:
 \begin{itemize}
 \item{Step 1: \emph{CP Violation} } \\
Utilising $C$, $S$, $ \mathcal{A}_{\Delta \Gamma}$ and their CP conjugates, we determine $\xi$ and $\bar{\xi}$, respectively, unambiguously from the data. The product $\xi \times \bar{\xi}$, generalised to include NP, allows a theoretically clean determination of $\gamma_{\rm eff}\equiv 
 \gamma + \gamma_{\text{NP}}$, where $\gamma_{\rm NP}$ is a function of the NP parameters. Using information on $\gamma$ from other processes, we extract $\gamma_{\rm NP}$. 
\item{Step 2: \emph{Branching Ratio Information}} \\
To have a particularly clean setup, we combine the branching fractions of the non-leptonic decays with differential rate information from their semi-leptonic partners. We constrain the exchange topologies via other control channels. Complementing data with theoretical input for the $|a_1|$ parameters, we extract $b$ and $\bar{b}$.  \item{Step 3: \emph{NP Parameters Correlation}} \\
We make use of all three observables $\gamma_{\rm eff}$, $b$ and $\bar{b}$ and explore the available space for NP by obtaining correlations between the NP parameters $\rho(\varphi)$ and $\bar{\rho}(\bar{\varphi})$. 
 \end{itemize} 
%
%
%
\section{Concluding Remarks}
 \begin{figure}[t!]
	\centering
	\includegraphics[width = 0.74\linewidth]{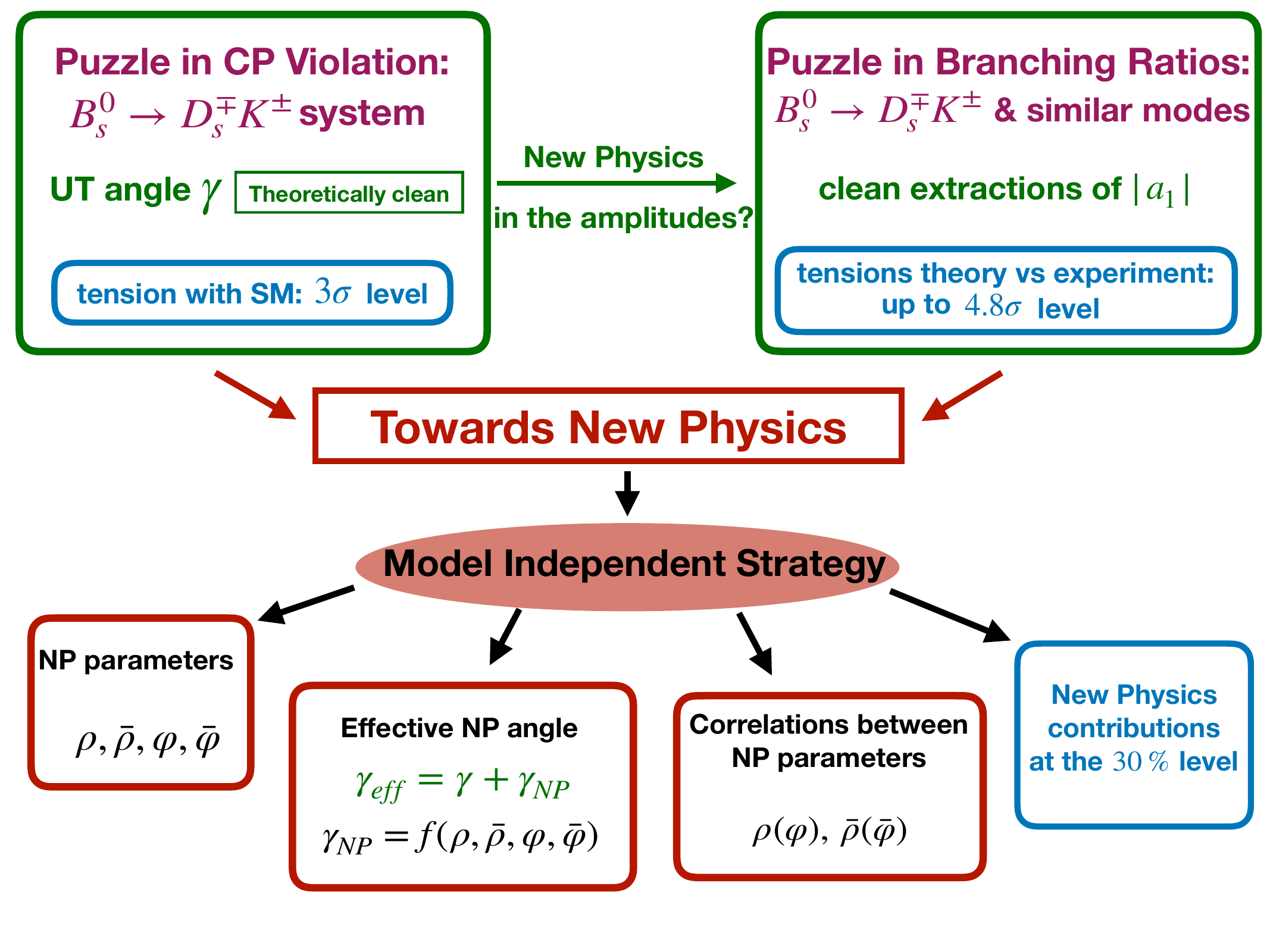}
	\caption{Illustration of the two puzzles and the strategy we presented in this paper.} \label{fig:flow-chart}
\end{figure}
The $\bar{B}^0_s\to D_s^+K^-$ and $\bar{B}^0_s\to D_s^- K^+$ decays with their CP conjugates are key players in the testing of the SM. We have demonstrated that the intriguing picture arising from the measured CP-violating observables, which results in a tension with the SM at the $3\sigma$ level, is complemented by a puzzling pattern of the individual branching ratios. The latter finding is actually in accordance with measurements of rates of 
$B_{(s)}$ modes with similar dynamics, where we find tensions with up to $4.8\sigma$ significance. While the $\gamma$ measurement cannot be explained through non-factorizable effects, the branching ratio puzzles could in principle be accommodated through such contributions. We would like to stress that the experimental result of the strong phase difference $\delta_s$ between the $b\to c \bar u s$ and $b\to u \bar c s$ decay paths is in excellent agreement with factorization. It is exciting to reveal and link these new puzzles, complementing indications of NP in other corners of the flavour sector, where currently rare decays arising from $b\to s\ell^+\ell^-$ quark-level processes involving also leptons are in the spotlight \cite{Albrecht:2021tul}.

We have presented a model-independent description of the $B^0_s\to D_s^\mp K^\pm$ system to include NP effects and give a generalized expression for an effective angle $\gamma$ in terms of NP parameters that can be determined in a theoretically clean way from the measured CP-violating observables. Utilizing furthermore the information from the decay branching ratios, the NP parameters can be constrained, as we have illustrated for the current data. In 
Fig.~\ref{fig:flow-chart}, we give a schematic overview. In the future high-precision $B$ physics era, this formalism can be fully exploited through more sophisticated experimental analyses, searching for direct CP violation and measuring the experimental $B_s$ branching ratios for the separate $D_s^+K^-$ and $D_s^- K^+$ final states. The measurement of the differential $\bar B^0_s \rightarrow K^+ \ell \bar{\nu}_{\ell}$ rate would be another important ingredient to complement the analysis. The NP parameters resulting from this analysis will serve as benchmarks for the model building community.

It will be exciting to monitor the evolution of the data and the sharper picture emerging from the application of our strategy. 
The central question is whether these studies will finally allow us to establish the presence of new CP-violating contributions to $B^0_s\to D_s^\mp K^\pm$ decays, thereby raising the question of whether such new sources of CP violation 
could open a window to understand the matter--antimatter asymmetry of the Universe.

%
%
%
\section*{Acknowledgements} 
We would like to thank Ruben Jaarsma, Philine van Vliet and Kristof De Bruyn for useful discussions. 
This research has been supported by the Netherlands Organisation for Scientific Research (NWO).


%
%
%

%
%
%
\end{document}